\documentclass[%
 reprint,
superscriptaddress,
 amsmath,amssymb,
 aps,
]{revtex4-1}

\usepackage[pdftex]{graphicx}
\usepackage{graphicx}
\usepackage{dcolumn}
\usepackage{bm}
\usepackage{lineno}
\usepackage{amsmath}
\usepackage{amsfonts}

\usepackage{xr-hyper}
\usepackage{graphicx}
\usepackage{amsmath,amssymb}
\usepackage{bm}
\usepackage{hyperref}
\usepackage{xcolor}
\usepackage{setspace}

\hypersetup{
    colorlinks=true,
    linkcolor=blue,
    citecolor=blue,
    urlcolor=blue
}

\begin{document}

\title{Photon density of states engineering with generative inverse design for scalable 3D photonic metamaterials}

\author{
Zesen Zhou$^{1,2,*}$,
Jeevan Rois$^{1,2}$,
Matias Kagias$^{1,2,*}$ \\[1em]
{\footnotesize
$^{1}$Division of Synchrotron Radiation Research and NanoLund, Department of Physics, Lund University, Sweden\\
$^{2}$Wallenberg Initiative Materials Science for Sustainability, Department of Physics, Lund University, Sweden
}
}

\email{zesen.zhou@fysik.lu.se}
\email{matias.kagias@fysik.lu.se}

\begin{abstract}
The photon density of states (pDOS) governs fundamental light–matter interactions and is a critical parameter for designing next generation light driven technologies such as photocatalysis and solar energy harvesting. Achieving a target pDOS in three-dimensional (3D) nanoarchitected structures remains challenging due to the nonlinear and non-unique relationship between geometry and spectral response. Here, we present an end-to-end inverse design framework for tailoring the pDOS of 3D photonic metamaterials fabricated via the scalable nanofabrication approach of metasurface-based holographic lithography. A data-driven forward surrogate model is constructed to predict frequency-resolved pDOS spectra from metasurface diffraction parameters and lithographic thresholds. Inverse design is performed using a conditional generative adversarial network (cGAN) that generates candidate metasurface diffraction parameters for target pDOS features.  3D structures featuring high local pDOS were obtained across a broad normalized frequency range and consistently outperformed those in the original dataset. Structural analysis revealed that these high-pDOS architectures fall into two predominant structural categories with similar rotational symmetry characteristics. Our work establishes the first inverse design strategy for 3D photonic metamaterials fabricated via holographic lithography.
\end{abstract}

\maketitle

\begin{figure*}[hbt!]
    \centering
    \includegraphics[width=0.8 \textwidth]{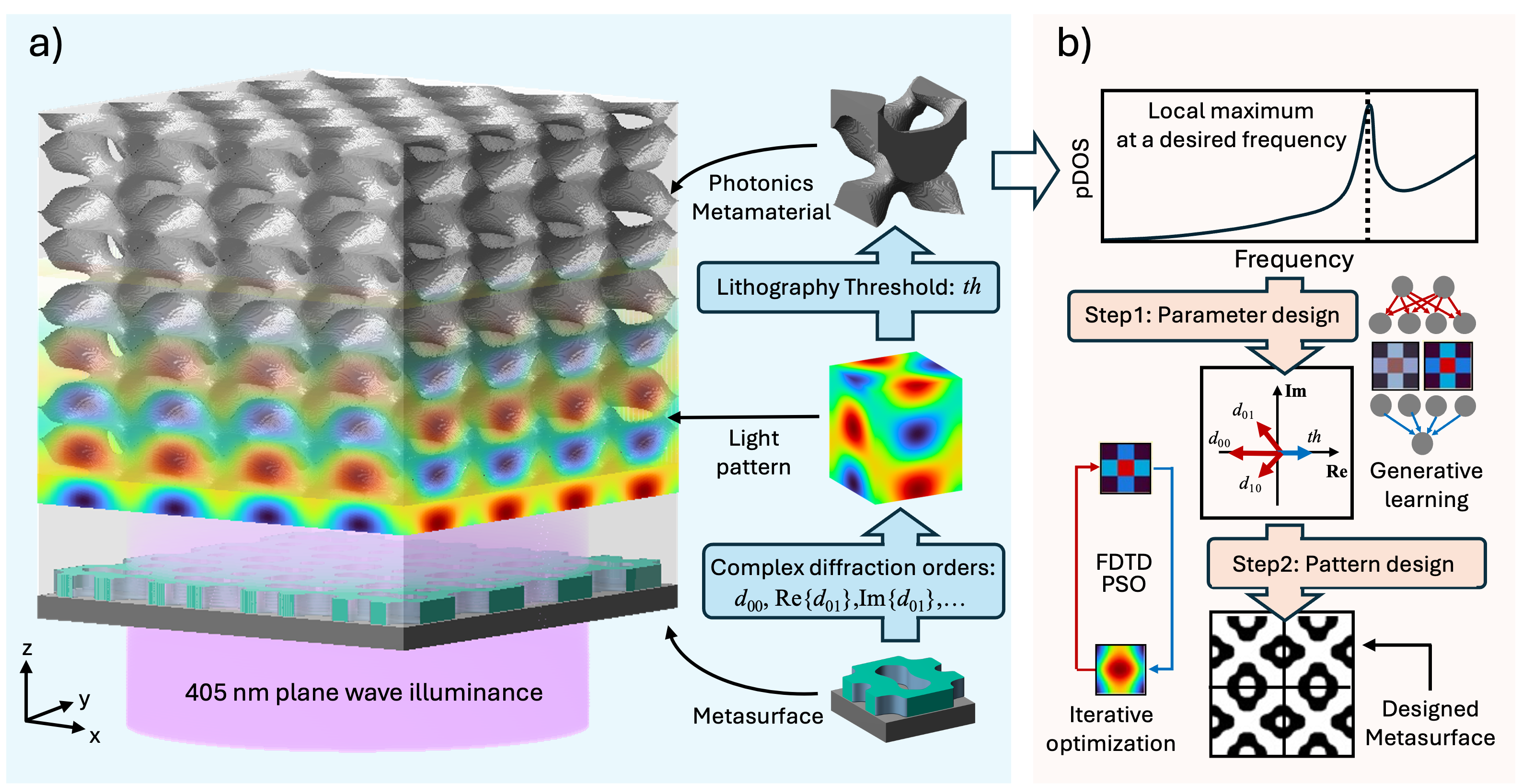}
    \caption{\textbf{Metasurface-based holographic lithography and inverse design.}
    (a) The metasurface generates complex diffraction orders under 405 nm plane-wave illumination, which interfere to form a 3D periodic light pattern. Applying a lithographic threshold converts the intensity distribution into a nano-architected photonic structure, whose geometry determines the resulting pDOS characteristics.
    (b) End-to-end inverse design framework linking prescribed pDOS features to a fabricable metasurface. Diffraction parameters are first generated to target desired spectral responses, followed by metasurface pattern retrieval through FDTD-assisted particle swarm optimization.}
    \label{fig:concept}
\end{figure*}

\section*{Introduction}
The photon density of states (pDOS) quantifies the number of available electromagnetic modes per unit volume and frequency in a photonic medium \cite{Yablonovitch1987}. According to Fermi’s golden rule, the spontaneous emission rate of a quantum emitter is proportional to the local density of states (LDOS), which is directly determined by the underlying pDOS \cite{Lodahl2004}. Control over the pDOS thus provides a direct route to engineering emission rates and light–matter interactions \cite{Jiang2011,Lodahl2015}. This can be extremely useful for light energy based technologies such as radiative cooling, photocatalysis, and solar energy harvesting \cite{Zhao2023, wang2025, Sun2025, Baranov2019, Shi2018}.

.

Photonic crystals and metamaterials provide a platform for tailoring the pDOS through periodic modulation of the refractive index distribution \cite{McPhedran2004, Fussell2004, Johnson2002}. This periodic modulation gives rise to photonic band structures governed by Bloch modes, where the redistribution of electromagnetic states in reciprocal space leads to frequency regions with suppressed or enhanced pDOS \cite{Atkin1996}. The total number of electromagnetic states remains conserved in accordance with the Barnett–Loudon sum rule, so spectral suppression in one frequency range is necessarily accompanied by enhancement elsewhere \cite{Barnett1996}. The first laboratory realization of 3D photonic metamaterial in 1991, known as \textit{Yablonovite}, demonstrated that periodic dielectric architectures can support a complete photonic band structure \cite{YABLONOVITCH1991}. Subsequent advances in nanofabrication (such as colloidal self-assembly \cite{Norris2001}, interference lithography \cite{Campbell2000}, and two-photon direct laser writing \cite{Serbin2004}) have enabled a wide variety of architectures, including inverse opals \cite{Varghese2013}, diamond-like lattices \cite{Galusha2008}, woodpile \cite{Willenswaard2024}, and more complex morphologies such as gyroid minimal-surface networks \cite{Dolan2015, Peng2016} and disordered hyperuniform structures \cite{Siedentop2024}. In most of the presented cases, the achievable pDOS tuning is largely confined to adjustments of lattice constant or filling fraction, ignoring structural complexity and morphology as control parameters. The non-intuitive and non-linear relationship between spectral properties and 3D morphology suggests that fine pDOS tuning is achievable via fine morphological design \cite{Cersonsky2021, Cersonsky2026}. Delivering optimized photonic architectures that can be translated into light based technologies for sustainability, requires both manufacturing methods with a high design freedom and scalability as well as inverse design frameworks to derive fabricate-able and robust morphologies.

Metasurface-enabled interference lithography can produce complex 3D photonic architectures with both high structural complexity and scalability \cite{Kondo2006, Kamali2019}. Metasurfaces can generate near field interference light patterns (Fig. \ref{fig:concept} (a)) that are used for photolithographic fabrication \cite{Lee2022,Kagias2023}. The high controllability originates from the unique wavefront shaping capabilities of dielectric metasurface \cite{Zhang2020}. The metasurface determines the intensity, phase, polarization, and propagation angles of the diffraction modes that interfere to achieve the desired 3D architectures. Although some inverse design works have explored the relationship between 3D morphology and its optical response \cite{Chen2025,Meng2018,Wang2022,Men2014,Cersonsky2026}, no work so far has attempted end to end design approach linking metasurface design with final performance of the 3D morphology.

In this work, we develop an inverse design framework for tailoring the pDOS of 3D photonic metamaterials generated by metasurface-based holographic lithography (Fig. \ref{fig:concept} (b)). We first establish a forward prediction model using a long short-term memory (LSTM) network that learns the mapping between the metasurface diffraction parameters, the exposure threshold, and the resulting pDOS spectrum. The trained model accurately predicts the pDOS response and serves as a fast surrogate model for inverse design. The inverse design then aims to retrieve metasurface parameters from prescribed pDOS features. We employ a two-step architecture in which conditional generative adversarial networks (cGANs) generate candidate diffraction parameters and exposure thresholds corresponding to the target pDOS characteristics, after which structurally connected candidates are selected and used in a coupled finite difference time domain (FDTD) and particle swarm optimization (PSO) co-simulation to reconstruct the metasurface patterns. The designed 3D photonic architectures exhibit pronounced enhancement of the normalized pDOS at prescribed target frequencies and maintain strong performance across a broad frequency range. We further identify two representative types of 3D morphologies that effectively enhance the pDOS at the target frequency. Our framework establishes metasurface-based holographic lithography as a programmable platform for pDOS engineering of scalable photonic architectures.

\section*{Results}

\noindent \textbf{Forward prediction of photon density of states (pDOS)} For the forward prediction model we assume a metasurface with a periodicity of 500 nm that is illuminated by a 405 nm plane wave, producing five propagating diffraction orders \cite{Hua2012} annotated by the zeroth-order mode $(0,0)$ and the first-order modes $(\pm1,0)$ and $(0,\pm1)$ (Fig. \ref{fig:concept} (a)). These propagating diffraction orders interfere with each other and form a triply periodic light field with a body-centered-tetragonal (BCT) Bravais lattice \cite{Yuan2016}. By applying an intensity threshold to the interference field and taking its negative, the corresponding 3D architectures are obtained. Under the assumption that the generated structures are composed of a high-contrast material ($\varepsilon_{r} = 13$), the pDOS is calculated (see methods).

\begin{figure*}[htbp]
    \centering
    \includegraphics[width= 0.8 \textwidth]{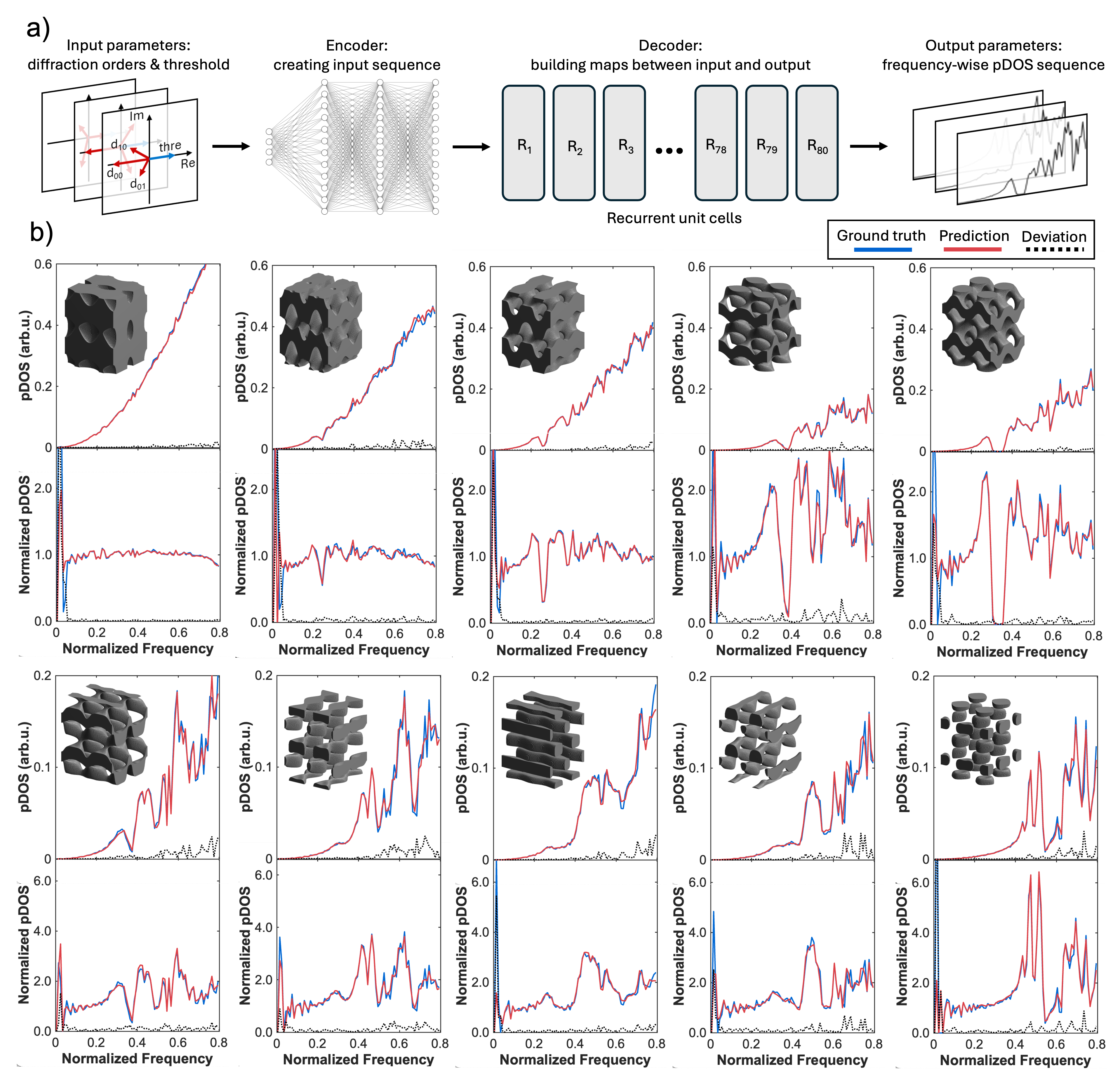}
    \caption{\textbf{Forward prediction architecture and examples.}
    (a) Encoder--decoder architecture used for pDOS forward prediction. 
    The input parameters consist of the complex diffraction orders and the lithography threshold, 
    and the output is the frequency-resolved pDOS sequence.
    (b) Predicted results of pDOS and normalized pDOS. Blue, red, and black curves represent the ground truth,
    prediction, and deviation, respectively. The upper panels show the pDOS spectra of the generated
    3D architectures with the corresponding morphologies shown in the insets. The lower panels show
    the pDOS normalized by the long wavelength approximation $\propto \omega^{2}$.}
    \label{fig:forward_model}
\end{figure*}

Based on this process, we numerically generate a dataset consisting of 45,000 samples. For each sample, the real and imaginary parts of the first-order diffraction coefficients, $\mathrm{Re}(d_{0,\pm1})$, $\mathrm{Im}(d_{0,\pm1})$, $\mathrm{Re}(d_{\pm1,0})$, and $\mathrm{Im}(d_{\pm1,0})$, together with the zeroth-order coefficient $d_{0,0}$ and the threshold value $t_h$ are taken as inputs. The output is the corresponding pDOS consisting of 80 discrete points. Statistical analysis of the peaks and band gaps of the generated dataset confirms that high-pDOS samples occupy only a small fraction of the parameter space, highlighting the necessity of data-driven inverse design approaches (Fig. S1).

The forward prediction model (Fig. \ref{fig:forward_model} (a)) is based on an encoder–decoder architecture implemented with LSTM networks. The encoder first transforms the input parameters into a sequence representation to match the dimensionality of the output pDOS sequence, while the decoder captures the nonlinear relationships between the input parameters and the resulting pDOS response. This architecture enables efficient learning of the global dependencies within the pDOS sequence while maintaining consistency between the input parameters and the predicted spectral response (see methods). 


\begin{figure*}[htbp]
    \centering
    \includegraphics[width= 0.8 \textwidth]{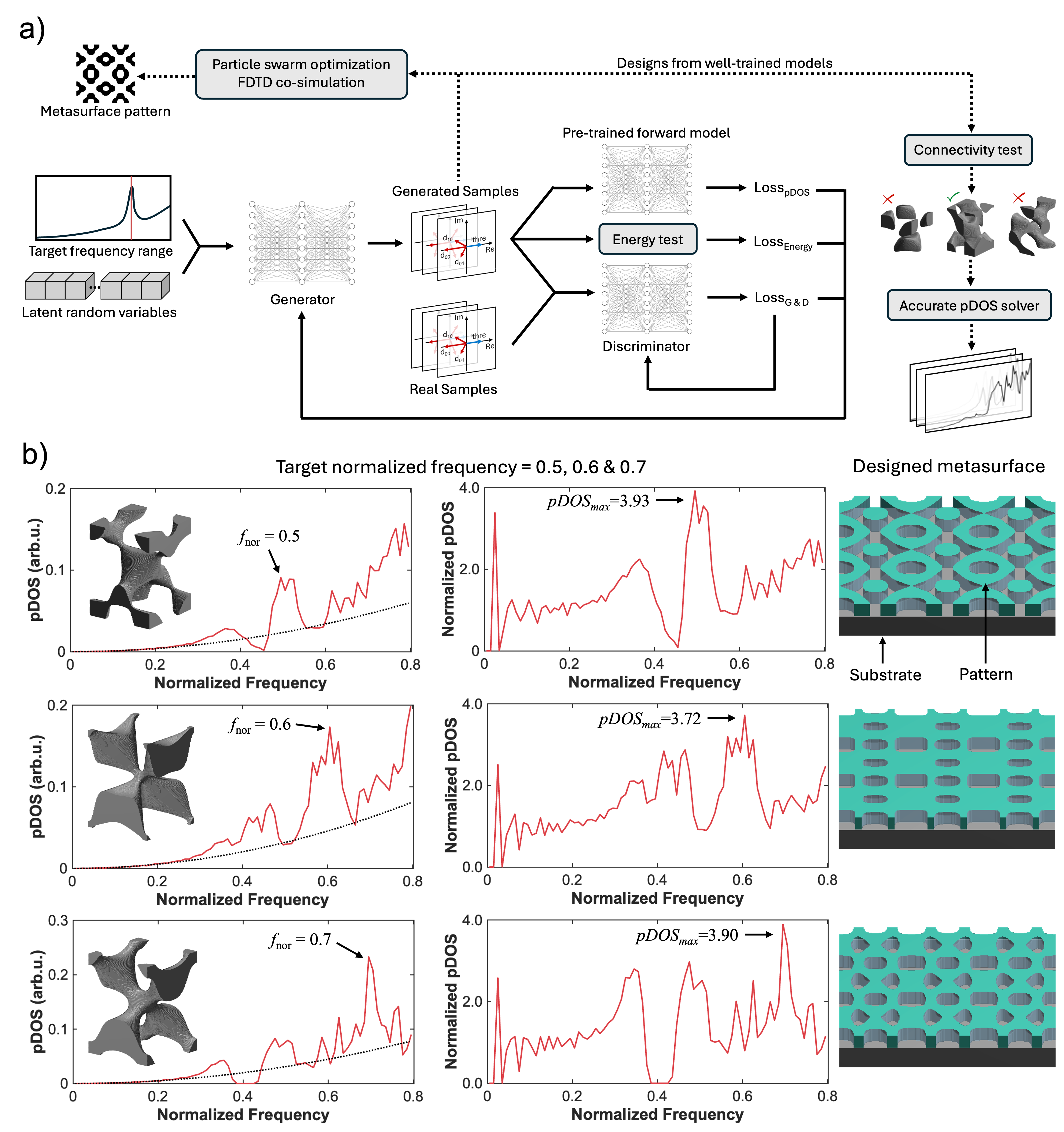}
    \caption{\textbf{Inverse design architecture and examples.}
    (a) Architecture of the inverse design framework. A conditional GAN generates candidate diffraction orders and threshold values from the target frequency range and latent variables. The generated samples are evaluated using the pre-trained forward model and energy conservation constraint, followed by a connectivity test. Valid candidates are then converted into fabricable metasurface patterns through FDTD-assisted particle swarm optimization (PSO).
    (b) Representative inverse-designed structures targeting normalized frequencies of 0.5, 0.6, and 0.7. The left panels show the corresponding 3D morphologies and pDOS spectra, while the middle panels present the normalized pDOS highlighting the enhanced peaks near the target frequencies. The right panels show the reconstructed metasurface patterns.}
    \label{fig:metsurface}
\end{figure*}
Selected forward prediction examples with different pDOS features (Fig. \ref{fig:forward_model} (b))
showed a strong agreement between ground truth and predictions. The model captures the overall spectral trends as well as several local peaks and minima with high accuracy. The errors remain low in the low-frequency region (0.2--0.5), where the maximum deviation is below $\sim$0.1, while larger deviations appear at higher frequencies (0.5--0.8), with the maximum error reaching  $\sim$0.5. These discrepancies arise from slight mismatches in peak intensities and small shifts in peak positions in reproducing local spectral features, but they have negligible impact on the subsequent inverse design as the surrogate model still provides reliable gradients for optimization. Additionally most 3D morphologies exhibiting strong pDOS enhancement features lose structural connectivity under particular combinations of diffraction orders and threshold values. This highlights the significance of utilizing generative models in the inverse design of physically meaningful structures.

\noindent \textbf{Inverse design of metasurface patterns} The core goal of our study is to inversely design metasurface patterns given the desired local pDOS features. The implemented inverse design framework contains two main steps (Fig. \ref{fig:metsurface} (a)). In the first step, a cGAN generates candidate diffraction coefficients and threshold values from the target frequency range and latent random variables. The generator is optimized using a joint loss function consisting of an adversarial loss, a pDOS-target loss, and an energy-conservation loss (see methods). This combination enables the model to efficiently generate samples that match the target pDOS while enforcing physically consistent diffraction intensities, and the pDOS-target loss further drives the generation toward enhanced pDOS responses through gradient-based optimization. In the second step, the generated samples are filtered through a connectivity test to ensure structurally connected architectures. The corresponding metasurface patterns are then reconstructed through finite difference time domain (FDTD) and particle swarm optimization (PSO)
co-simulation, completing the inverse design process. 

\begin{figure*}[htbp]
    \centering
    \includegraphics[width=0.6\textwidth]{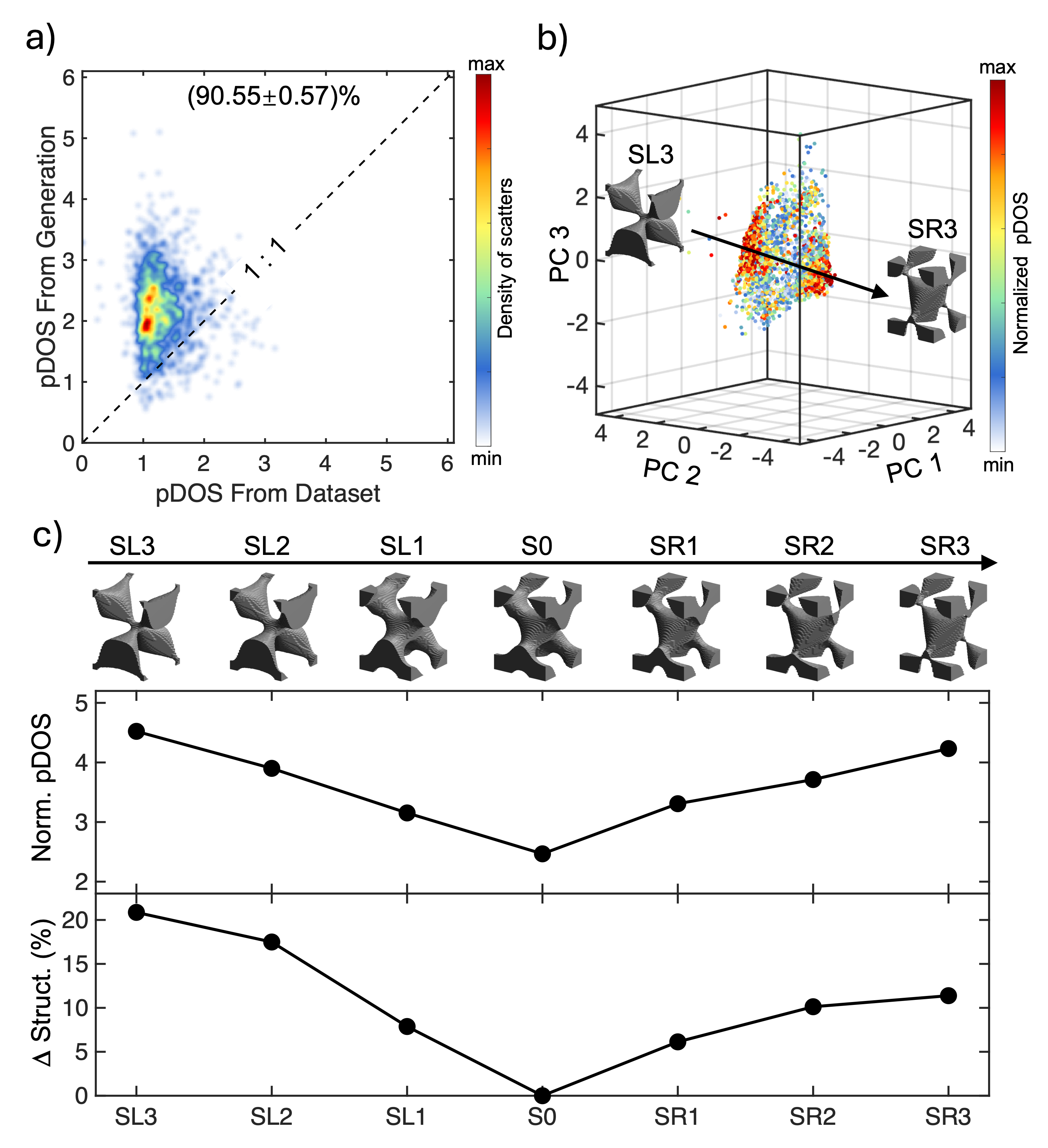}
    \caption{\textbf{Structural characteristics of GAN-generated samples.}
    (a) Density map of the scatter distribution comparing of normalized pDOS values between GAN-generated samples (vertical axis) and dataset samples (horizontal axis). The color scale indicates the density of sample pairs in the scatter distribution. Regions above the dashed $y=x$ line indicate cases where the GAN sample yields a higher performance than the dataset sample. 
    (b) Low-dimensional representation of the generated structures obtained by autoencoder compression followed by principal component analysis (PCA). Each point represents one generated structure in the reduced feature space, while the color scale indicates the normalized pDOS value, revealing two dominant structural families..
    (c) Continuous transition of the normalized pDOS value and structural morphology between the two representative families.}
    \label{fig:figure4}
\end{figure*}

The cGAN model is trained within a broad normalized frequency range from 0.45 to 0.70. A total of 10,000 samples are generated, from which top-ranked candidates that also satisfy the structural connectivity constraint are selected for further analysis. To inverse-design structures with locally enhanced pDOS, a normalized pDOS value larger than 2 at the target frequency is used as the threshold to identify qualified samples. Under this criterion, an average generation fidelity of approximately 80\% is achieved across the entire frequency range, indicating that a large fraction of the generated samples fall within the desired design space.

Example inverse design cases for peak normalized pDOS at the three normalized frequencies of 0.5, 0.6, and 0.7 showed broad variability both in 3D morphology and end metasurface design (Fig. \ref{fig:metsurface} (b)). Owing to the large design freedom enabled by the irregular binary patterns, the resulting diffraction orders match the target diffraction distributions with high precision, exhibiting an average deviation below 0.001. These results confirm that the inverse-designed pDOS features can be translated into physically realizable metasurface geometries. The corresponding 3D morphologies of the designed structures, exhibit similar symmetry characteristics. In the \(xy\) plane, the architectures display clear \(C_2\) symmetry, originating from the symmetric distribution of the diffraction coefficients at \((\pm1,0)\) and \((0,\pm1)\) (see methods). Along the \(z\) direction, the upper and lower halves of the structures form approximately rotated counterparts with a rotation of about \(90^\circ\), giving rise to a characteristic twisted morphology. Further end-to-end design results exhibit similar behavior (Fig. S4).

The normalized frequency step size is set to 0.01 here, and its influence on the observed normalized pDOS enhancement is further examined by varying the step size from 0.001 to 0.04 (Fig. S5). All three cases exhibit good robustness, as the peak normalized pDOS remains above 3 when the frequency step size is reduced, indicating that the designed structures achieve pronounced and robust pDOS enhancement near the target frequencies.

A second cGAN was trained to inverse-design 3D structures exhibiting broadband vanishing pDOS, corresponding to photonic band gaps (Fig. S6). In contrast to the pDOS-enhancement cases, the formation of bandgaps is found to depend more strongly on the exposure threshold which is equivalent to the volume fraction (VF) rather that fine morphological variations \cite{Ho1990}. 

\vspace{1pc}

\noindent \textbf{Morphological variation in inverse-designed samples} To further evaluate the performance of the cGAN-generated high-pDOS structures, we use the density distribution to compare the normalized pDOS values of generated samples with those of the dataset samples (Fig. \ref{fig:figure4} (a)). For each target-frequency group, 500 generated samples are paired with randomly selected dataset samples for direct comparison. Most data points (90.55\%) lie above the 1:1 reference line, indicating that the generated structures generally exhibit higher normalized pDOS than the corresponding dataset samples. The small fraction of points below the equality line can be attributed to the finite generative accuracy of the model as well as the presence of a limited number of high-pDOS samples already contained in the dataset. Notably, while the best-performing dataset samples reach normalized pDOS values of around 3, the GAN-generated structures achieve values approaching 5, demonstrating that the generative model is capable of exploring regions of the design space with significantly enhanced pDOS.

Analysis of the 3D morphology is performed using an autoencoder-based principal component analysis (PCA) method to reduce dimensionality (Fig. S7). The autoencoder effectively transforms the 3D structural data into three-dimensional latent-space variables while preserving most of the structural information. PCA is then applied to the latent space variables to further analyze and rank these latent variables. The distribution of the principal components (PC1, PC2, and PC3) of the latent space variables showed two distinct branches (Fig. \ref{fig:figure4} (b)). Moving from the center toward the two sides of the distribution, the structures exhibit progressively higher pDOS values. The structures located on the two sides display similar rotational symmetries but different geometric morphologies. These observations indicate that structures exhibiting high pDOS are confined to specific regions of the structural parameter space, and that the cGAN-based learning framework is capable of effectively identifying these regions. As the structures evolve continuously from one type (labeled as SL3--SL1) to the other (labeled as SR1--SR3) (Fig. \ref{fig:figure4} (c)), the normalized pDOS at the target frequency first decreases from nearly 5 to about 2.5 and then increases again to approximately 4.5. During this transition, the structural parameters vary (labeled as $\Delta$ Struct.) by about 20\% and 12\% along the two sides of the distribution, respectively. This behavior reveals a trade-off in realizing these morphologies. The reference structures located near the center exhibit thicker connecting branches, leading to more mechanically robust networks but relatively lower pDOS values. In contrast, the structures on the two sides (SL3 and SR3) possess much thinner connecting features, which contribute to stronger pDOS enhancement but may become mechanically fragile and more prone to collapse during fabrication. On the other hand, practical fabrication processes may introduce imperfections from both the metasurface pattern and the holographic lithography process. The results indicate that despite these structural variations, the designed architectures are able to maintain relatively high pDOS values, suggesting that the inverse-designed structures exhibit a certain degree of robustness against fabrication errors.

\begin{figure}[htbp]
    \centering
    \includegraphics[width=0.45\textwidth]{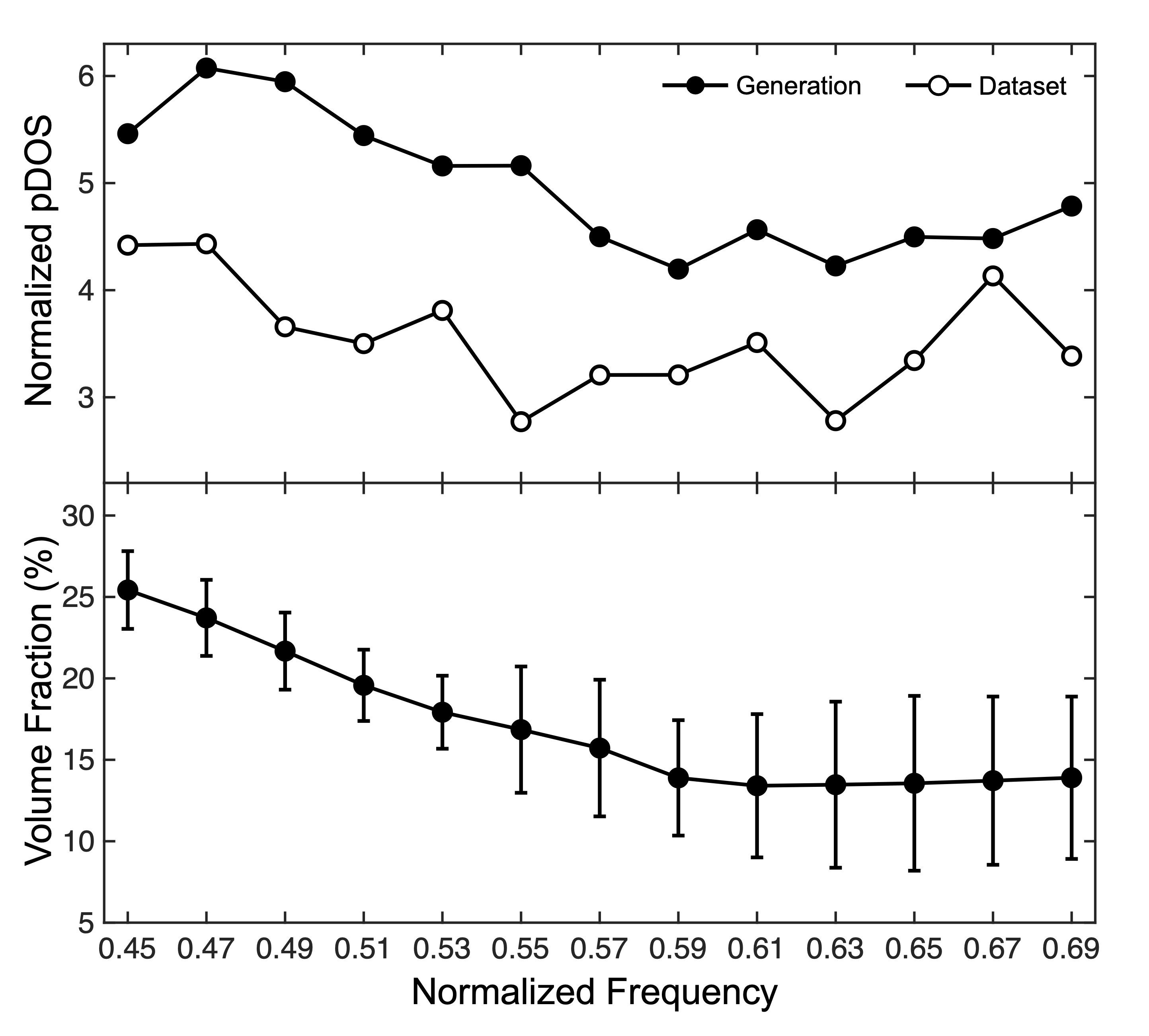}
    \caption{\textbf{Frequency-dependent performance of GAN-generated samples.}
    Comparison of the pDOS enhancement achieved by GAN-generated samples and dataset samples across the studied normalized frequency range from 0.45 to 0.69 and corresponding volume fraction distribution.}
    \label{fig:figure5}
\end{figure}

\enlargethispage{\baselineskip}

The cGAN consistently reaches higher maximum pDOS values over a broad spectral interval, demonstrating its ability to explore high-performing regions of the design space (Fig. \ref{fig:figure5}). In particular, the generated structures exhibit normalized pDOS values above 4.2 across all investigated frequency intervals, and in some cases approach values close to 6. In contrast, the highest pDOS values observed in the dataset remain limited to approximately 2.7--4.1. This improvement arises because the GAN captures the characteristic feature distribution of high-pDOS structures in the dataset and is further guided by the pretrained forward model through the local pDOS loss formulation (see methods), which drives the generation toward structures with enhanced pDOS responses. To validate this interpretation, additional ablation analyses were performed (Fig. S8), where the performances of the GAN alone and the local pDOS loss alone were evaluated separately. The results show that using the GAN alone leads to inferior performance and using only the local pDOS loss fails to achieve stable convergence. Regarding the average VF of the cGAN-generated structures as a function of the target normalized frequency, a decreasing trend is observed from approximately 25\% to 12\% as the frequency increases from 0.45 to 0.61. At higher frequencies, however, the average VF remains nearly constant while the fluctuation, indicated by the standard deviation, becomes noticeably larger. This behavior reflects different mechanisms governing the pDOS enhancement across the spectral range. In the lower-frequency region, the enhancement mainly occurs near the band edges associated with photonic band gaps (red and blue edges), whose frequencies scale inversely with the volume fraction of the structure \cite{Ho1990}. In contrast, at higher frequencies the enhanced pDOS features are mainly related to higher-order photonic modes, which impose weaker constraints on the overall volume fraction and therefore lead to larger structural variability.

\section*{Conclusions}
An end-to-end inverse design framework for tailoring the photon density of states (pDOS) in holographically fabricated three-dimensional (3D) photonic metamaterials has been developed. 3D structures exhibiting locally enhanced pDOS are realized through specific combinations of diffraction orders and exposure thresholds generated by a conditional generative adversarial network (cGAN), and these diffraction parameters are further reconstructed into metasurface patterns via FDTD-assisted particle swarm optimization (FDTD-PSO). Quantitative comparison with the original dataset shows that the inverse-designed high-pDOS structures consistently achieve substantially enhanced responses across a broad frequency range and maintaining structural diversity. Further structural analysis reveals that those elaborate 3D architectures fall into two distinct morphological categories while sharing similar rotational symmetry characteristics. This indicates an underlying structural characteristic in realizing high pDOS. On the one hand, stronger pDOS enhancement is associated with thinner connecting features, which increases fabrication difficulty. On the other hand, the overall pDOS response remains relatively robust against moderate geometric deviations, suggesting tolerance to fabrication induced imperfections. Our work provides a framework for on demand pDOS design in three-dimensional architectures and clarifies the structure property relationship, enabling the directed realization of functional 3D photonic structures via holographic lithography.

\section*{Methods}

\noindent \textbf{Dataset generation} The dataset is constructed to establish a quantitative mapping between metasurface-defined diffraction parameters and the resulting pDOS of 3D photonic structures. The input features consist of the complex diffraction coefficients together with the photolithography exposure threshold that determines the final 3D structure morphology. Five independent parameters describe the diffraction components: the zeroth-order diffraction coefficient $A_{0,0}$, chosen as the phase reference, and the real and imaginary parts of the first diffraction orders along two orthogonal directions, denoted as $(\mathrm{Re}(d_{0,\pm1}),\mathrm{Im}(d_{0,\pm1}))$ and $(\mathrm{Re}(d_{\pm1,0}),\mathrm{Im}(d_{\pm1,0}))$. To ensure physically consistent interference fields, the zeroth-order amplitude is sampled within the range $0.5$--$1$, which is stronger than the first-order components range $0$--$0.5$, and all diffraction parameters are normalized such that the total optical energy equals unity.

For each parameter set, the diffraction orders are converted into the complex electric-field distribution on the metasurface plane using an inverse Fourier transform. The field is subsequently propagated along the axial direction using the angular spectrum method to generate a 3D interference intensity pattern \cite{Jiang2019}. Applying the exposure threshold $t_h$, randomly sampled within the range $0.1$--$0.5$, converts the interference field into a binary structure representing the fabricated structures. To account for volumetric shrinkage during interference lithography, all generated structures are scaled along the propagation direction by a factor of $0.7$ \cite{Denning2011}.

The pDOS of each reconstructed structure is then calculated using an in-house plane-wave expansion method (PWEM) code. Calculations have been demonstrated by replicating Ref.\cite{Florescu2005}. Approximately $1,800$ $k$-points are sampled within the irreducible Brillouin zone to obtain accurate eigenfrequencies. In total, $22,500$ independent samples are generated through this procedure. Data augmentation is further applied by exchanging the diffraction components $d_{0,1}$ and $d_{1,0}$, which yield identical pDOS responses due to structural symmetry, thereby expanding the dataset to $45,000$ samples. Each sample contains six input parameters and an output pDOS spectrum consisting of $80$ discrete frequency points within the normalized range $0$--$0.8$. The dataset is finally divided into training and testing subsets with a ratio of $4:1$. 

\vspace{1pc}

\noindent \textbf{Forward prediction model: LSTM} To predict the pDOS spectrum from the diffraction parameters, a neural network surrogate model based on LSTM networks is employed. The model adopts a standard encoder--decoder architecture based on LSTM units \cite{Deng2022}. The encoder first maps the input diffraction parameter vector to a latent representation through a fully connected layer with a Tanh activation, which initializes the hidden state of the LSTM decoder. A learnable sequence template with the same length as the output spectrum is then provided as the input of the LSTM decoder, enabling the network to generate the pDOS spectrum along the frequency axis. The decoder outputs are projected to scalar values at each step through a linear layer to form the predicted pDOS sequence.

The network uses an input dimension of $6$, a hidden dimension of $256$, and two stacked LSTM layers. The output sequence length is fixed to $80$ corresponding to the sampled frequency points of the pDOS spectrum. The model is trained by minimizing the mean squared error (MSE) between the predicted and ground-truth pDOS sequences using the Adam optimizer \cite{Kingma2015}. The initial learning rate is set to $10^{-3}$ with a step-decay learning-rate scheduler (step size $100$ and decay factor $0.8$). The batch size is $100$ and the network is trained for $2001$ epochs. Before training, both input parameters and pDOS outputs are normalized using min--max scaling. Training and testing performances are shown in Fig. S2(a)--(b).
\vspace{1pc}

\noindent \textbf{Inverse design model: cGAN} The inverse design model is implemented using a conditional generative adversarial network (cGAN), which generates diffraction parameters conditioned on the target frequency \cite{Goodfellow2014}. The generator takes as input a latent random vector $\mathbf{z}$ and the conditioning frequency $f$. These variables are concatenated and passed through a multilayer perceptron consisting of two hidden layers with 64 neurons and ReLU activations. The output layer produces six parameters corresponding to the exposure threshold and diffraction coefficients. The exposure threshold is constrained to the physically valid range of $0.1$--$0.5$ through a sigmoid transformation. The latent dimension is set to $8$. The discriminator receives both the generated structure parameters and the corresponding frequency condition and predicts whether the pair is real or generated. It consists of two fully connected layers with $64$ neurons and LeakyReLU activations followed by a sigmoid output layer \cite{Maas2013}.

The generator is trained using a composite loss composed of three terms:
\begin{equation}
\mathcal{L}_{\mathrm{G}}
=
\mathcal{L}_{\mathrm{adv}}
+
\lambda_{\mathrm{pDOS}} \mathcal{L}_{\mathrm{pDOS}}
+
\lambda_{\mathrm{E}} \mathcal{L}_{\mathrm{E}},
\end{equation}

The adversarial loss is defined as
\begin{equation}
\mathcal{L}_{\mathrm{adv}}
=
-
\mathbb{E}_{\mathbf{z},f}
\left[
\log D(G(\mathbf{z},f),f)
\right].
\end{equation}

The pDOS-target loss is defined as
\begin{equation}
\mathcal{L}_{\mathrm{pDOS}}
=
\sum_{f \in \Omega}
\left|
\mathrm{pDOS}_{\mathrm{pred}}(f)
-
\mathrm{pDOS}_{\mathrm{target}}(f)
\right|.
\end{equation}

The energy-conservation loss enforces normalization of the diffraction orders:
\begin{equation}
\mathcal{L}_{\mathrm{E}}
=
\left|
\sum_{m} |d_{m}|^{2} - 1
\right|.
\end{equation}

Training is performed using the Adam optimizer with learning rate $10^{-2}$ and batch size $1024$. Both generator and discriminator are trained simultaneously for $10{,}000$ iterations using a step-decay learning-rate scheduler with decay factor $0.9$ every $1000$ iterations. The weighting parameter for the energy-conservation term is set to $\lambda_{\mathrm{E}} = 1000$. Training and testing performances are shown in Fig. S2(c)--(d).

\vspace{1pc}

\noindent \textbf{Inverse design model: FDTD-PSO simulation} Metasurface patterns that reproduce the target diffraction orders are obtained using a co-simulation framework that combines particle swarm optimization (PSO) with full-wave FDTD simulations (Fig.~S3) \cite{Zhou2009}. The optimization process operates in an iterative loop as follows. The metasurface within a unit cell is parameterized in the Fourier domain by a set of coefficients defining its transmission function. For each particle in the PSO swarm, these Fourier coefficients are transformed into a real-space geometry function via an inverse Fourier transform. The resulting continuous geometry function is then binarized to generate a physically realizable metasurface pattern, which serves as the input geometry for the electromagnetic simulation.

Full-wave simulations are performed using Lumerical FDTD under normally incident illumination. The electric-field distribution on the metasurface plane is recorded using a field monitor and transformed into the Fourier domain to extract the simulated diffraction orders. These diffraction orders are compared with the target values obtained from the inverse design stage to evaluate the fitness of each particle. The PSO algorithm then updates the Fourier coefficients based on the particle velocities and historical best solutions. The swarm size is set to $50$ particles and the optimization runs for $30$ iterations with inertia weight $w=0.7$ and acceleration coefficients $c_1=c_2=1.5$. The design variables are constrained within the normalized range $[0,1]$.

\bibliographystyle{naturemag}
\bibliography{refs}

\section*{Data availability}
\noindent The data that support the findings of this study are available from the corresponding author upon reasonable request.

\section*{Acknowledgments}
\noindent This work has been funded by the Wallenberg Initiative Materials Science for Sustainability (WISE) supported by the Knut and Alice Wallenberg Foundation for financial support. The authors acknowledge Lund Nano Lab (LNL) for nanofabrication facilities. Computational resources were provided by the Swedish National Infrastructure for Computing (NAISS) and by the Center for Scientific and Technical Computing at Lund University (LUNARC).

\section*{Author contributions}
\noindent ZZ designed and implemented the inverse design strategy, ran the simulations, and conducted the analysis. JR fabricated the metasrufaces. MK conceived and lead the project. ZZ wrote the manuscript with input from all authors.  

\section*{Additional Information}
\noindent Supplementary information is available and includes: S1, statistical analysis of the dataset; S2, training and validation results of neural networks; S3, the FDTD-PSO simulation framework; S4, additional end-to-end inverse design results; S5, analysis of the dependence of pDOS enhancement on frequency sampling; S6, inverse design of band gap structures; S7, Autoencoder-based principal component analysis (PCA); and S8, ablation studies of the generative model.

\clearpage

\end{document}